\documentstyle[aps]{revtex}
\begin{document}
\draft
\title{Resonant Magnetization Tunneling in Mn$_{12}$ Acetate: The Absence of
Inhomogeneous Hyperfine Broadening}
\author{Jonathan R. Friedman}
\address{Department of Physics and Astronomy, The State University of New York at\\
Stony Brook, Stony Brook, NY 11794-3800}
\author{M. P. Sarachik}
\address{Physics Department, City College of the City University of New York, New\\
York, New York 10031}
\author{R. Ziolo}
\address{Wilson Center for Research and Technology, Xerox Corporation, Webster, New\\
York 14580}
\date{\today}
\maketitle

\begin{abstract}
We present the results of a detailed study of the
thermally-assisted-resonant-tunneling relaxation rate of Mn$_{12}$ acetate
as a function of an external, longitudinal magnetic field and find that the
data can be fit extremely well to a Lorentzian function. No hint of
inhomogeneous broadening is found, even though some is expected from the Mn
nuclear hyperfine interaction. This inconsistency implies that the tunneling
mechanism cannot be described simply in terms of a random hyperfine field.
\end{abstract}


\smallskip The molecular magnet Mn$_{12}$ acetate, with spin 10 and large
uniaxial anisotropy, has garnered much attention since its hysteresis loop
was found to exhibit steps at regular intervals of magnetic field\cite
{fried1}. This phenomenon, now confirmed by other experiments on this \cite
{hernan1,thomas,lionti,friedjap,luis1} and other materials \cite
{aubin1,sangregorio,aubin2,ruiz}, has been interpreted as a manifestation of
resonant tunneling of the magnetization, first suggested to occur in this
system in Refs. 11 and 12. The system is modeled as a
double-well potential (Fig. 1) with energy levels that correspond to the
different projections of the spin along the easy axis. An external magnetic
field will tilt the potential. At specific values of field, levels in opposite
wells come into resonance and thermally assisted tunneling between the wells
becomes possible (solid arrows in Fig. 1), increasing the interwell
relaxation rate and thereby producing steps in the hysteresis loops. The
dashed arrows in Fig. 1 schematically illustrate the nonresonant process of
simple thermal activation over the classical energy barrier (which in some
cases may differ from the full barrier shown\cite{fried2}).

Mn$_{12}$ acetate has been quantitatively described\cite
{fried1,thomas,sessoli} by the Hamiltonian: 
\begin{equation}
{\cal {H}}=-DS_z^2-g\mu _BH_zS_z+{\cal {H^{\prime }}}
\end{equation}
where $D\approx 0.6K$ represents the anisotropy energy that breaks the
21-fold zero-field Zeeman degeneracy and ${\cal {H^{\prime }}}$ is a
perturbation that does not commute with $S_z$. (Recent EPR experiments\cite
{barra,hill} have indicated the presence of additional terms in the
Hamiltonian that are fourth order in the spin operators.) In the absence of
a symmetry-breaking term, ${\cal {H^{\prime }}}$, $S_z$ is conserved and
hence no tunneling is allowed between levels. The symmetry of ${\cal {
H^{\prime }}}$ determines a selection rule that governs which level
crossings may give rise to tunneling. Many early theoretical treatments\cite
{enz,chudnovsky,hemmen} of tunneling in spin systems focused on a transverse
anisotropy of the form $kS_x^2$, which leads to a selection rule $\Delta
m=2i $ for integer $i$. It was then somewhat surprising that this selection
rule was violated\cite{fried1,thomas,hern} in the case of Mn$_{12}$. That
is, the rule implies that every other step should be missing. In fact, all
steps are observed and there is no discernible difference
between steps that obey the selection rule and those that violate it. This
prompted the suggestion\cite{hern,garaninandchud} that the tunneling must
instead be produced by a transverse magnetic field, ${\cal {H^{\prime }}}%
=g\mu _BS_xH_T$, which imposes no selection rule. Since no such field was
applied externally in the experiments, it was concluded that the transverse
component of an internal field of dipolar or hyperfine origin may be
responsible\cite{hern}. Dipolar fields have now largely been ruled out by
recent experiments\cite{caneschi}. On the other hand, hyperfine fields
within a Mn$_{12}$ cluster have been estimated to be on the order of a few
hundred Oe\cite{hartmann} and estimates\cite{fried2,hern,garanin} have shown
that this is sufficient for tunneling to occur between levels near the top
of the energy barrier, allowing the thermally assisted tunneling process
illustrated in Fig. 1.

Much recent theoretical activity \cite
{garaninandchud,burin,prokofev1,dobro,gunther,fort,luis2,prokofev2,garg} has
focused on determining the details of the relaxation process. Many of these
studies predict a resonance width that is much narrower than that observed,
a discrepancy that has been addressed by assuming that the resonances are
inhomogeneously broadened\cite{garaninandchud,prokofev1,dobro,luis2} by
dipolar or hyperfine fields or by misalignment of the crystallites in
oriented-powder samples. In this report, we present the results of detailed
measurements of the relaxation rate as a function of magnetic field in the
neighborhood of a resonance. We find that the data can be fit with high
fidelity to a Lorentzian function and show no hint of inhomogeneous
broadening. The linewidth corresponds to a time scale that does not
match any relevant microscopic time known for the system. We suggest that it
may represent a hitherto unrecognized component of the relaxation process.

A powder of Mn$_{12}$ prepared according to\cite{lis} was oriented in
a 5.5 Tesla field and set in a paraffin matrix in a manner similar to
the procedure described in Refs. 1 and 20.  Hysteresis loops for this
sample were published in reference 20; the squareness of the loops
indicates the crystallites were well oriented. Measurements were
performed with a Quantum Design MPMS-5 magnetometer; the
superconducting magnet was quenched prior to taking data to eliminate
any remanent field. We estimate the misalignment of a typical
crystallite to be $\approx 1$ degree. For the largest magnetic field
of 500 Oe applied in the present study, this misalignment corresponds
to a maximum unintentional transverse field of $\approx 9$ Oe, a value
much too small to have any effect on the tunneling, which is thought
to be produced by a transverse field of several hundred Oersteds.

Measurements were performed as follows. The sample was cooled in zero field
from 5 K (where the sample is superparamagnetic) to below the blocking
temperature of 3 K. After allowing the system to thermally stabilize, a
magnetic field was applied and the magnetization was measured as a function
of time to obtain its relaxation rate. We note\cite{hern,fried3} that the
field experienced by a molecular cluster is the superposition of the
externally applied field and the internal mean dipolar field produced by the
other clusters. Compared with measuring the relaxation from a finite
magnetization toward M $\approx $ 0, the above procedure provides the
advantage that the internal field associated with the change in the sample
magnetization changes little compared to the externally applied field, so
that the total field is almost constant.

A typical relaxation curve is shown on a semilogarithmic scale in the inset
to Fig. 2. As noted in several previous publications \cite
{fried1,hern,paulsen1,paulsen2}, the single-exponential relaxation expected
for a sample ostensibly comprising identical molecules is not observed. We
suggest that this may be due to slight variations in the local environment of
the clusters, perhaps related to the number of proximal molecules of
solvation. To eliminate the effect of such variations, we measured the
longest relaxation time by fitting the long-time tail of the relaxation to a
single exponential, as shown in the inset to Fig. 2. The high quality of the
fit indicates that all faster processes have died out. The relaxation rate
obtained this way at 2.6 K is plotted as a function of applied field in the
main part of Fig. 2.\cite{note1} The qualitative signature of resonant
tunneling is apparent: when the field increases from zero, the relaxation
rate decreases as matching levels in opposite wells become detuned. The
figure also shows the results of fitting the data to a Lorentzian and to a
Gaussian, each superimposed on a background of the form $\Gamma
_{nonres}Cosh(g\mu _BSH/k_BT)$, which represents the non-resonant,
over-barrier part of the relaxation.\cite{note2} The Lorentzian fit is
excellent while the Gaussian fit clearly is not, failing especially in the
tails.

On a semilogarithmic scale, Fig. 3 shows data taken at four different
temperatures, as indicated. A Lorentzian fit is applied to each set. From
these fits the full width is found to be $267,236,270$ and $271$ Oe for
2.5, 2.6, 2.7 and 2.8 K, respectively. This is consistent with the width
found by fitting ac susceptibility data to a Lorentzian\cite{luis1}. The
present experiments were limited to a small temperature range in which a
large portion of the relaxation could be measured in a reasonable time.
Because of this limitation, the effective energy barrier (an Arrhenius plot)
could not be deduced from the data, as has been done in ac measurements\cite
{luis1,novak,novak2}. However, one can get a crude estimate of the
difference in the height of the energy barrier on resonance and off
resonance if one assumes that in both cases the relaxation obeys
an Arrhenius law, $\Gamma =\omega _0e^{-U/kT}$, with a prefactor $\omega _0$
that is approximately independent of field. Then, the change in the
effective energy barrier, $\Delta U$, is given by $\Delta U=Tln(\Gamma
(H=0)/\Gamma _{nonres})$. The fits then yield $\Delta U=3.58,4.73,5.00$ and $
5.40$ K for $T=2.5,2.6,2.7$ and 2.8 K, respectively.

The absence of any apparent inhomogeneous broadening of the resonance
is quite unexpected. Since all resonances appear to have the same
height,\cite{fried1,thomas,hern} the tunneling must in large part be
driven by a transverse magnetic field of at least a few hundred
Oe.\cite{hern} The most likely source for such a field is the Mn
nuclear spins. This has been estimated\cite {hartmann} to be $300-500$
Oe. We stress that even were the hyperfine fields to play no role in
the tunneling, their presence is nevertheless expected and should give
rise to inhomogeneous broadening of the resonance. Luis et
al.\cite{luis1} have suggested that the broadening they observe is due
at least in part to interactions with hyperfine and dipole fields. We
note, however, that such interactions should give rise to a Gaussian
lineshape. It is puzzling that no such broadening is observed.

There are now numerous theories of relaxation in Mn$_{12}$, but none are
able to provide accurate, quantitative descriptions of the resonance
lineshape and linewidth. Dobrovitski and Zvezdin\cite{dobro} have considered
a model of tunneling out of the ground state that predicts a resonance
width several orders of magnitude lower than that observed. They suggest
that the larger observed widths are associated with inhomogeneous broadening
due to random dipolar or hyperfine fields. We note that the discrepancy may
in part be due to the fact that the relaxation is thermally assisted at
temperatures as low as 750 mK and there is evidence for
temperature-dependent relaxation even as low as 60 mK\cite{perenboom};
therefore, it seems unlikely that the observed relaxation (especially the
present data) can be described in terms of ground-state tunneling. In a
calculation that does not incorporate thermal activation, Gunther\cite
{gunther} has calculated the width of the hysteresis steps and concludes
that a dynamical transverse magnetic field must be invoked to account for
the discrepancy between his theory and experiment. Prokof'ev and Stamp\cite
{prokofev1} have argued that the dynamics of hyperfine and dipolar fields
must play a role in the relaxation and explicitly calculated how the
hyperfine interaction should give rise to Gaussian broadening of the
resonance. (In more recent work\cite{prokofev2}, the same authors offer a
calculation of the relaxation at short times and very low temperatures;
those results are not relevant to the present study.)

Garanin and Chudnovsky\cite{garaninandchud} (see also Friedman\cite{fried3})
have treated the relaxation of Mn$_{12}$ using a model of thermally assisted
tunneling in which the tunneling takes place from a level near the top of
the barrier and is produced by a static transverse magnetic field. They
predict the resonance to be a superposition of Lorentzians, and also invoke
inhomogeneous broadening by random fields to account for the broad
resonances observed. Fort {\it et al.}\cite{fort} have offered a calculation
of the resonance lineshape that fits the data for the zero-field resonance
reasonably well. Their calculation is based on the assumption that the
tunneling is driven by a fourth-order transverse anisotropy. As the authors
note, this approach fails to account for the presence of half of the observed 
resonances. Luis {\it et al.}\cite{luis2} considered a model in which the
tunneling is driven by both a transverse anisotropy and a transverse field.
This allows all the resonances, although their theory does predict some
difference in amplitude between the odd and even resonances that has not
been observed. Here, inhomogeneous broadening is invoked to smooth the
multiresonant results into a single peak\cite{fernan2}.

Some theories have attempted to explain the relaxation of Mn$_{12}$ in
terms of mechanisms other than tunneling. Burin, Prokof'ev and
Stamp\cite{burin} have suggested that the relaxation can occur via
dipolar flip-flop processes. This possibility has now been obviated by
measurements that show that the resonant phenomenon is substantially
unchanged when the Mn$_{12}$ molecules are dispersed in a glassy
matrix and thereby have negligible dipole
interactions\cite{caneschi}. Very recently, Garg\cite{garg} has
suggested that the relaxation may be due to a lattice distortion that
occurs when levels near the top of the barrier are near
resonance. This theory implies a correlation between the width of the
resonance and its height and therefore, Garg concludes, the
effective barrier is reduced on resonance by about 10 mK; in contrast,
experiments indicate that the barrier is reduced by several Kelvin
(see above and also \cite{luis1,novak2}).

How to interpret the observed lineshape and width remains an open question.
One tempting, but unlikely, interpretation is that the lineshape represents
the bare (coherent) tunneling rate. If this were the case, then the
linewidth would be a sensitive function of applied transverse field, an
effect that is not borne out by experiments\cite{friedjap}. Another
interpretation is that our results reflect the natural lineshape of the
levels that are involved in the tunneling, the width then being a measure of
the levels' lifetimes. If we assume that the tunneling is occurring between,
say, levels $m=3$ and $m=-3$, then the observed width of $\approx 250$ Oe
corresponds to a lifetime $h/g\mu _B(2m)H$ of $2.5$ x $10^{-10}s$. While
this is typical of spin-lattice relaxation times in many magnetic systems,
we note that the measured Arrhenius prefactor $\tau _0(=2\pi /\omega _0)$
for Mn$_{12}$ is around $10^{-7}s$ and it is this number that is expected
\cite{villain} to characterize the typical lifetime of excited states in the
system. We note this value for $\tau _0$ is anomalously large for
superparamagnetic systems, a fact that has been addressed theoretically\cite
{villain}. It seems, then, that there are three time scales involved in the
relaxation process: $\tau _0$, the tunneling time and the time scale
corresponding to the resonance width, whatever its meaning.

To summarize, the data presented here leave two mysteries: (i) Why, even in
the presence of significant hyperfine fields, is there no apparent
inhomogeneous broadening of the resonances?; and (ii) What is the origin of
the observed Lorentzian lineshapes? We conjecture that these questions may
be related and may be resolved simultaneously. At this point it is clear
that the role of the hyperfine interactions in the relaxation of Mn$_{12}$
cannot simply be understood in terms of a random field superimposed on the
external field. A more sophisticated model of the interaction of the
molecular spin with the nuclear spins seems essential to understand the
relaxation mechanism for Mn$_{12}$.

We thank Philip Stamp for some useful remarks about his work. Partial
support for this work was provided by the U.S. Air Force Office of
Scientific Research under grant F49620-92-J-0190 and the National Science
Foundation under grant DMR-9704309.

\begin{figure}[p]
\caption{Double-well potential for the Mn$_{12}$ Acetate, described by Eq. 1
with H = 0. The levels represent different projections of the spin along the
z axis (different values of the magnetic quantum number {\it m}). For an
initial excess population in the left well, the model of thermally assisted
resonant tunneling is represented by the solid arrows, while the nonresonant
process of simple thermal activation over the classical barrier is
represented by the dashed arrows.}
\label{fig1}
\end{figure}

\begin{figure}[p]
\caption{Relaxation rate as a function of applied magnetic field for the
zero-field resonance at 2.6 K. Fits to both Lorentzian and Gaussian
functions (with the same number of free parameters) are shown. The inset
shows the actual magnetization versus time (on a semilogarithmic scale) at
100 Oe. The long-time tail of the relaxation was fit to an exponential
(dashed line) to determine the relaxation rate for this field value.}
\label{fig2}
\end{figure}

\begin{figure}[p]
\caption{Semilogarithmic plot of the relaxation rate versus magnetic field
for the zero-field resonance at four different temperatures, as indicated.
Each data set was fit to a Lorentzian, as shown.}
\label{fig3}
\end{figure}


\begin{references}
\bibitem{note1}  The fits of the long-time tail of the data to an
exponential yield negligible statistical errors. The largest identifiable
random error in the data presented in Figs 2 and 3 is due to a $\approx 5$
Oe uncertainty in the field produced by the superconducting magnet. However,
we note there is a large error in the absolute value of the relaxation rate:
If we ignore the long-time tail of the relaxation and fit a limited region
of the earlier part of the data to an exponential, then the resulting rates
can be as much as 20 \% larger than those presented. Rates obtained in such
a way also fit a Lorentzian function extremely well and so our conclusions
remain unchanged.
\bibitem{note2}  This form for the nonresonant relaxation rate is valid for $
H<<H_c$, where $H_c\approx 10$ T is the critical field at which the barrier
disappears. $\Gamma _{nonres}$, the nonresonant relaxation rate at H = 0, is
the only fitting parameter in the expression.
\end{references}
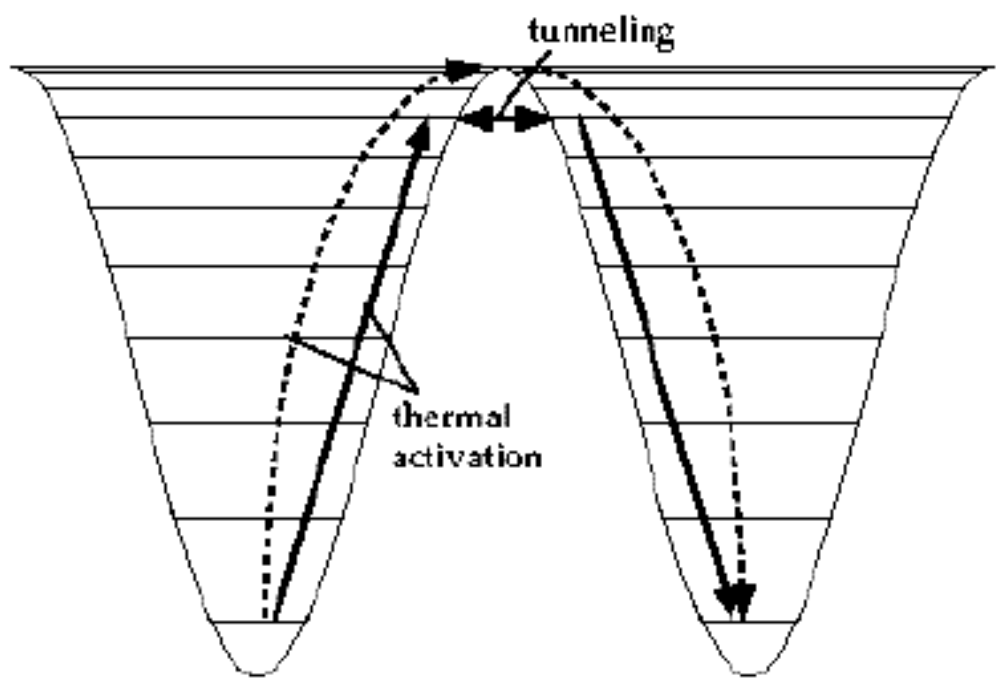
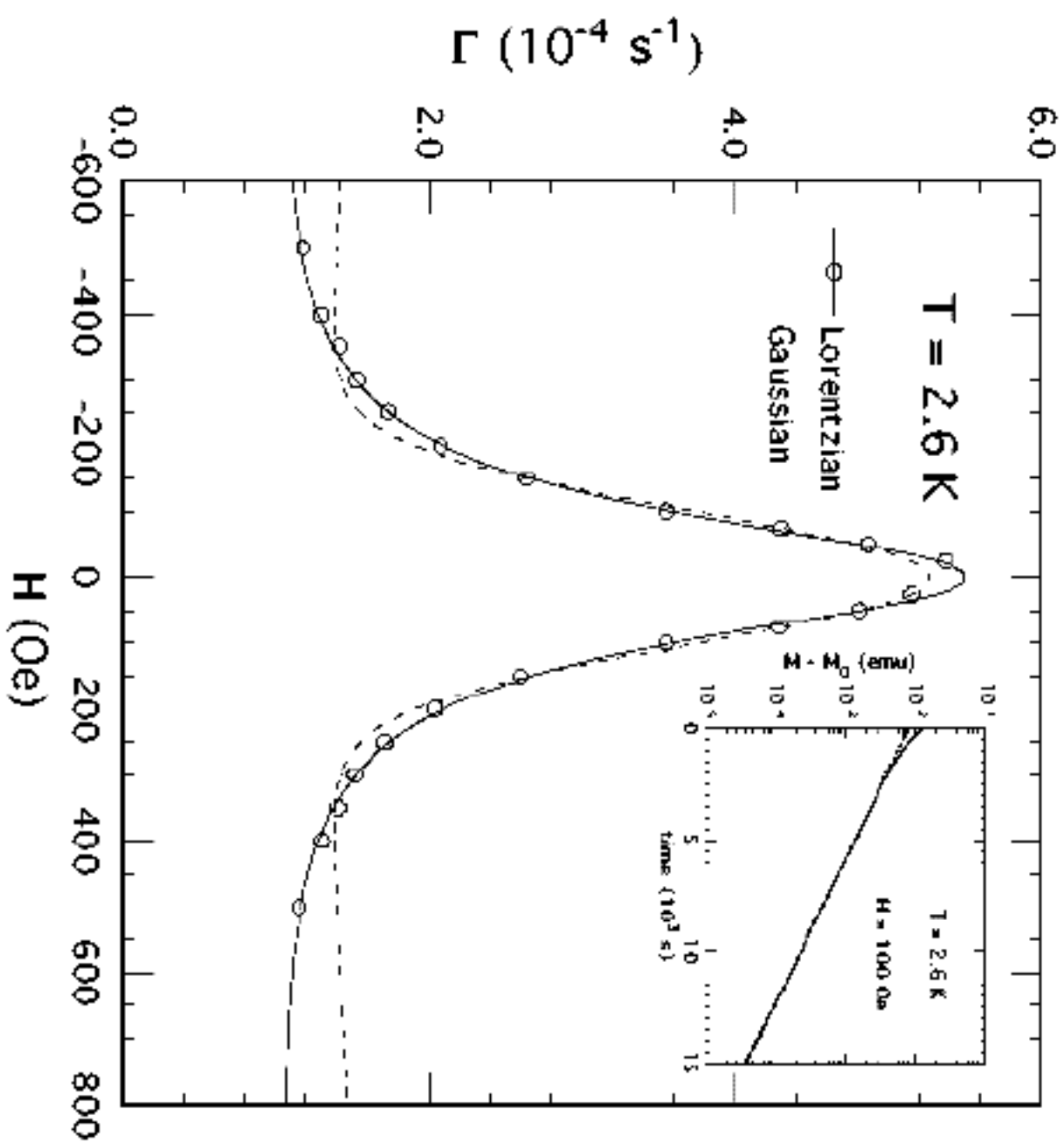
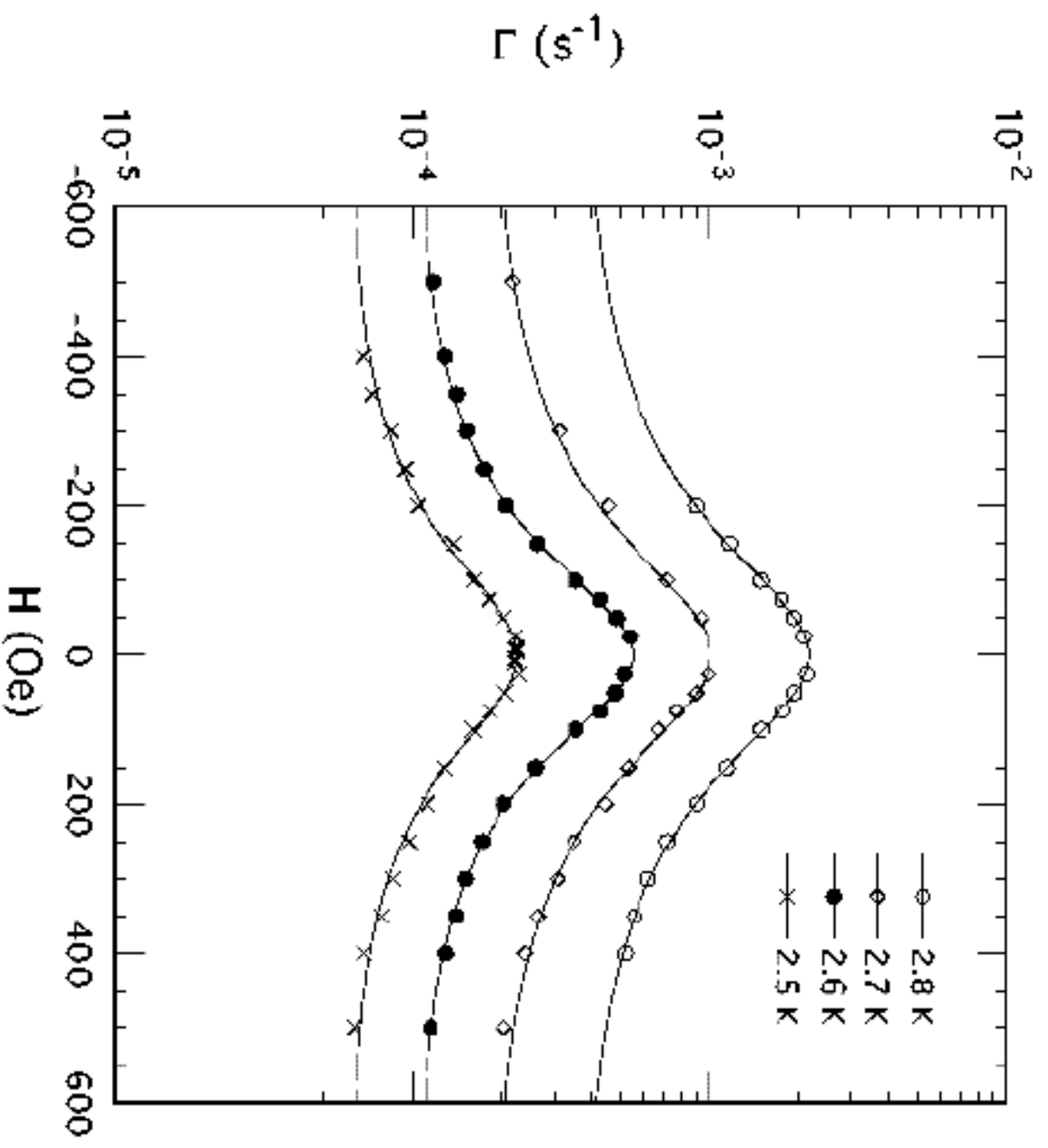
\end{document}